\documentclass[a4paper]{mem}
\usepackage{natbib}
\usepackage{graphicx}
\usepackage[a4paper]{hyperref}
\idline{00}{1}
\begin{document}
\title{Prospects for X-ray Studies of Galaxy Clusters with Astro-E2/XRS}

\author{T. Furusho \inst{1}, K. Mitsuda \inst{1}, N. Yamasaki\inst{1}, 
	R. Fujimoto \inst{1}, \and  T. Ohashi \inst{2}\fnmsep}

\offprints{T. Furusho}
\mail{3-1-1 Yoshinodai, Sagamihara, Kanagawa 229-8510, Japan}

\institute{Institute of Space and Astronautical Science (ISAS),
 Japan Aerospace Exploration Agency (JAXA),
 3-1-1 Yoshinodai, Sagamihara, Kanagawa 229-8510, Japan
 \email{furusho@astro.isas.jaxa.jp}\\ 
         \and
 Department of Physics, Tokyo Metropolitan University,
 1-1 Minami-Ohsawa, Hachioji, Tokyo 192-0397, Japan
             }

\abstract{ The Astro-E2 high resolution X-Ray Spectrometer (XRS) is
expected to provide a major enhancement in study of clusters of
galaxies. Astro-E2 is the fifth Japanese X-ray astronomy observatory,
which is scheduled for launch in early 2005. The XRS instrument,
developed under a Japan-US collaboration, is an X-ray microcalorimeter
with a capability of observing extended objects, and a high energy
resolution of about 6 eV at 6 keV. The spectral resolving power is 20
times higher than CCDs over the 0.5--10 keV energy band. We have
obtained several new results of clusters with Chandra and XMM, which
show that high-resolution imaging spectroscopy can clarify some
outstanding questions.  New sciences from Astro-E2 include the first
clear measurement of gas velocities, determination of ion and electron
temperatures, and electron densities based on the resolved line
features. We will describe the XRS instrument design, and present
simulations of the expected performance.
   }

   \authorrunning{T. Furusho et al.}
   \titlerunning{Prospects with Astro-E2/XRS}
   \maketitle
%

\section{Introduction}

Astro-E, the 5th Japanese X-ray astronomy satellite, was lost during
the launch on February 10, 2000 due to a malfunction with the 1st
stage of a M-V rocket. However, Astro-E was a very unique mission that
carried the first X-ray microcalorimeter into space and
would complement the great imaging performance of Chandra and the large
effective area of XMM-Newton. The rebuild mission Astro-E2 has been
approved, and is scheduled for launch in early 2005.

\begin{figure*} \centering
\includegraphics[width=0.56\textwidth]{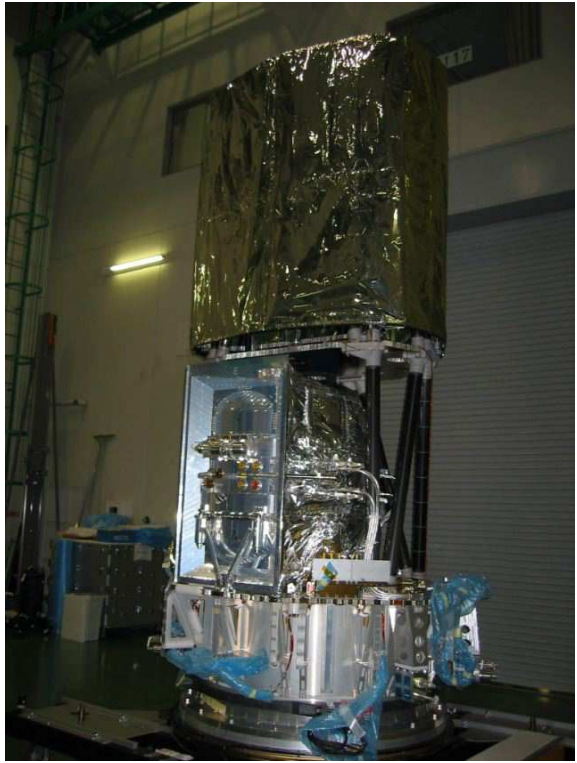}

\caption{The Astro-E2 satellite during the first spacecraft
integration test at ISAS/JAXA. In this photograph, the space craft was
not equipped with XRT or solar panels yet. The XRS Ne dewar
sits at the spacecraft base plate.}

 \label{E2photo} 
\end{figure*}

Figure \ref{E2photo} shows a photograph of the Astro-E2 spacecraft
under the first spacecraft integration test at ISAS/JAXA from July to
November, 2003. The satellite has three instruments: an X-ray
microcalorimeter (X-Ray Spectrometer; XRS), four X-ray CCD cameras
(X-ray Imaging Spectrometer; XIS), and a scintillation counter with
silicon PIN detectors (Hard X-ray Detector; HXD). XRS and XIS are put
at each focal plane of five X-ray thin foil mirrors (X-Ray Telescope;
XRT). They cover the soft energy band of 0.3--10 keV, while HXD
extends the bandpass of the mission up to 700 keV. The spacecraft is
approximately 1700 kg in weight, and about 6.5 m in length.
The XRS dewar occupies a large volume with a
weight of about 400 kg.  A M-V
rocket will bring the spacecraft into a circular near-earth orbit with
a nominal altitude of 550 km, and an inclination of 31$^{\circ}$.  The
detailed description of Astro-E2 and the progress are found at the web
site: {\tt http://www.astro.isas.jaxa.jp/astroe/ index.html.en}.

\section{X-ray spectrometer}

   \begin{figure*}
   \centering
   \resizebox{\hsize}{!}{
   \hspace*{4mm}\includegraphics[clip=true,width=0.46\textwidth]{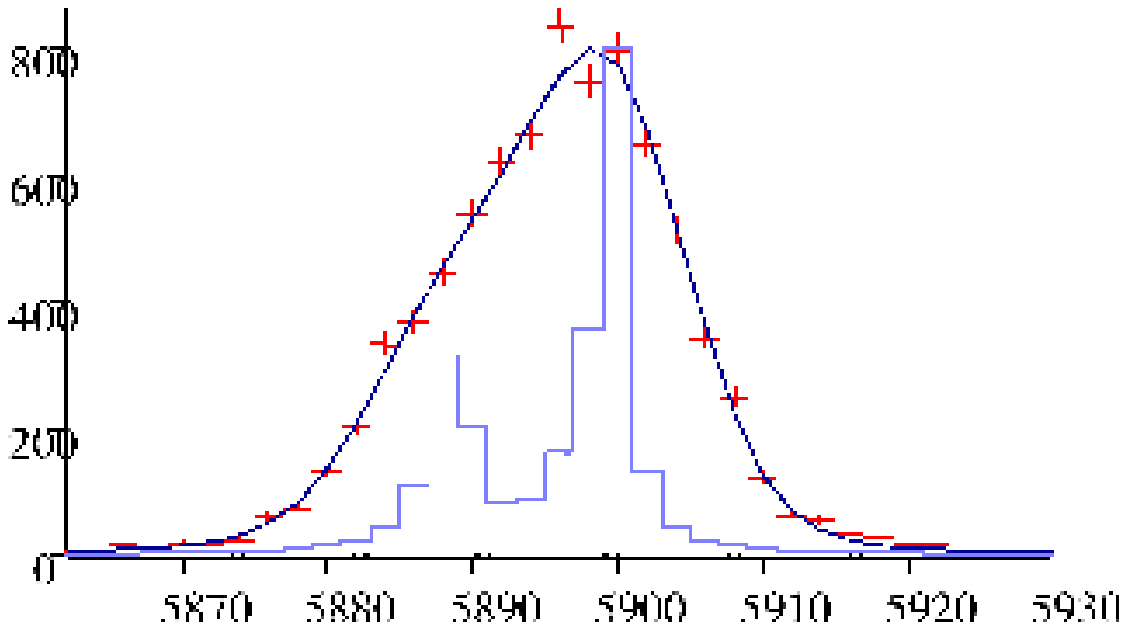}
   \hspace*{1mm}\includegraphics[clip=true,width=0.46\textwidth]{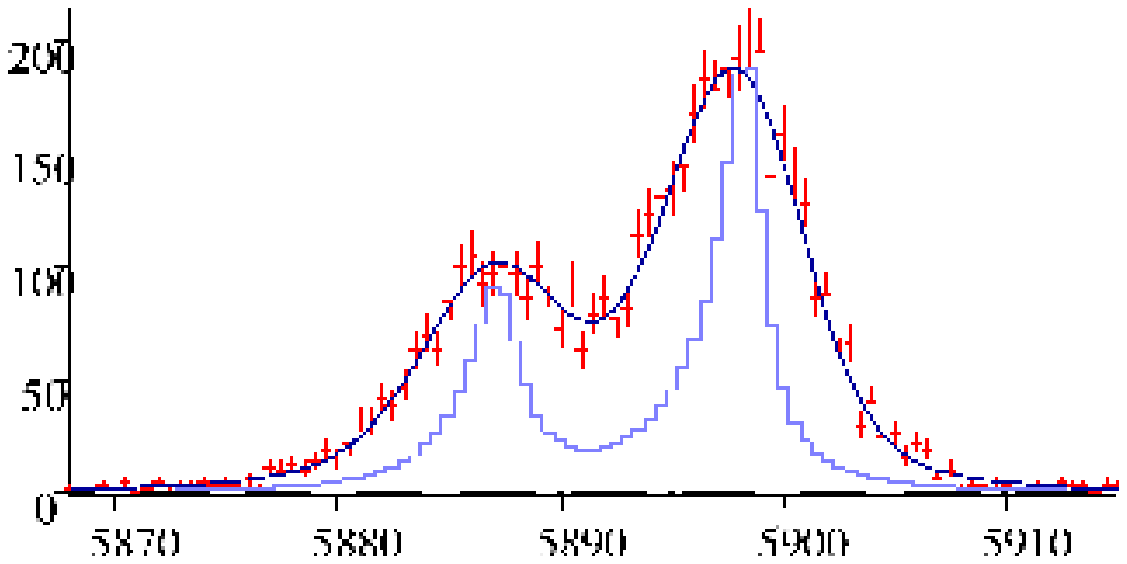}
   }
     \caption{Examples of Mn-Ka spectra taken by the original
bilinear array (left) and the new 2-dimensional array (right).}  
        \label{mnka}
    \end{figure*}
\begin{figure*} 
\centering
   \hspace*{2mm}
   \includegraphics[width=0.42\textwidth]{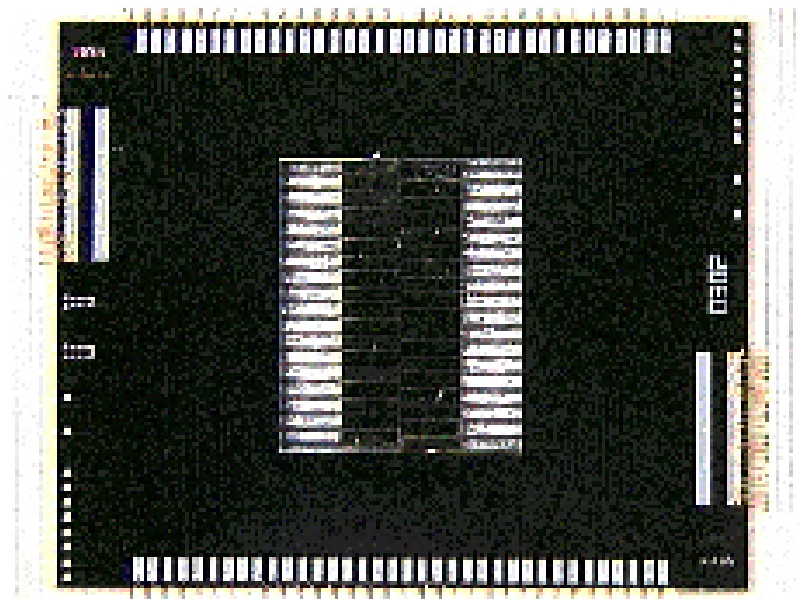}\hspace*{10mm}
   \includegraphics[width=0.37\textwidth]{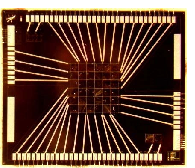}
   \caption{The original $2\times16$ bilinear array for
   Astro-E (left), and $6\times6$ 2-dimensional array for Astro-E2 (right).}
   \label{2dphoto}
\end{figure*}

XRS is the primary instrument of Astro-E2. The detector system was
developed at NASA/Goddard Space Flight Center and University of
Wisconsin. The baseline design of XRS for Astro-E2 is
same as the original XRS onboard Astro-E. The detector system consists
of 32 pixels with HgTe absorbers attached on silicon thermistors,
operates at a temperature of 60 mK with solid Ne, liquid He, and
an adiabatic demagnetization refrigerator to cool the detector down. 

\begin{table*} 
\centering
\caption{Performance of the original XRS of Astro-E and the new XRS of Astro-E2}
\hspace*{3mm}\resizebox{\hsize}{!}{
\begin{tabular}{lcc}
\hline\hline
            &  Astro-E & Astro-E2 \\\hline
Energy band &  0.3--10 keV & 0.3--10 keV\\
Energy resolution (FWHM at 6 keV) & 12 eV & 6 eV  \\
Field of view & $1.9'\times4.1'$ & $2.9'\times2.9'$\\
Array format & $2\times16$ & $6\times6$\\
Number of pixels & 32  &  31 +1(for calibration only)\\
Pixel size & $1.23\times0.318$ mm & $0.624\times0.624$ mm \\
Lifetime & $<2$ years & 2.5--3 years\\
\hline
\end{tabular}
}
\label{xrs}
\end{table*}

Once the Astro-E2 mission program was started, enormous efforts have
been continued to make improvements based on our precious knowledge of
the original XRS by practice. The XRS performance for both
Astro-E and Astro-E2 is summarized in Table \ref{xrs}. 
The following
items are most remarkable improvements done for Astro-E2/XRS.

{\it Higher energy resolution:} Figure \ref{mnka} shows examples
of $^{55}$Fe spectra (Mn-K lines) obtained from the original and
new XRS \citep{caroline}. The new XRS has an energy resolution of 6 eV
at 6 keV ($E/\Delta E=1000$), a factor of 2 better than the
original. The high spectral capability enables us to resolve
Mn-K$\alpha_1$ and K$\alpha_2$ lines beautifully, of which difference
of the peak energies is 11 eV.

{\it Longer lifetime:} A major design change of the cooling system is an
addition of a mechanical cooler outside the Ne dewar to improve a
lifetime of XRS longer. The lifetime of the original XRS was limited
to about 2 years by the amount of liquid He. With the mechanical
cooler, the lifetime is expected to be at least 2.5 yr, or more.

{\it Lower background:} We also have changed a placement of
in-flight calibration sources of $^{55}$Fe and $^{41}$Ca, which were
originally attached inside the detector system that caused a
significant background. The calibration sources will be attached to
a filter wheel, which is located above the XRS outside the Ne
dewar. Users can choose a filter with them, or without them. Instead
of the internal calibration sources, we dedicate one of the 32 pixels
for calibration only that is illuminated by a $^{55}$Fe source. The
design change of the calibration sources results in lower internal
background and much higher sensitivity of the new XRS.

{\it Better detector design:} We adopt a 6$\times$6 2-dimensional
array covering $2.9'\times 2.9'$ field of view as shown in the right
panel of Figure \ref{2dphoto}, instead of a 2$\times$16 bilinear array
as shown in the left panel.  The 2-dimension array is more
suitable for observations of celestial objects as it results in
lots of benefits of higher energy resolution, faster pulse response,
etc.

Those improvements make XRS even more fascinating system. Now the
flight XRS array is under the ground calibration tests, and we will
start the final spacecraft integration tests in April, 2004.

\section{Feasibility study for science with galaxy clusters}

\begin{figure} [h]
\centering
\hspace*{-2mm}\includegraphics[clip=true,angle=-90,width=7.1cm]{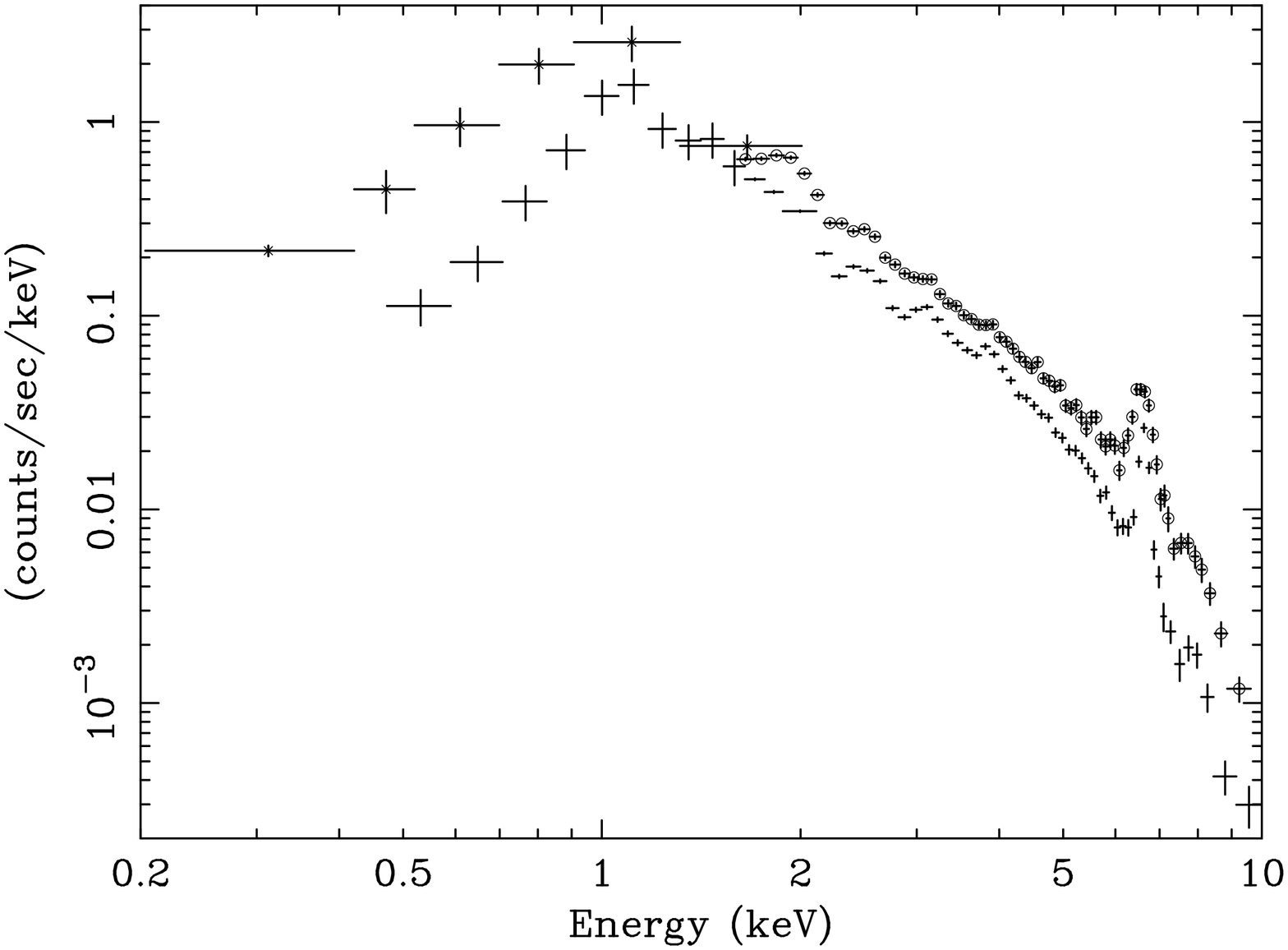}
\includegraphics[clip=true,angle=-90,width=6.6cm]{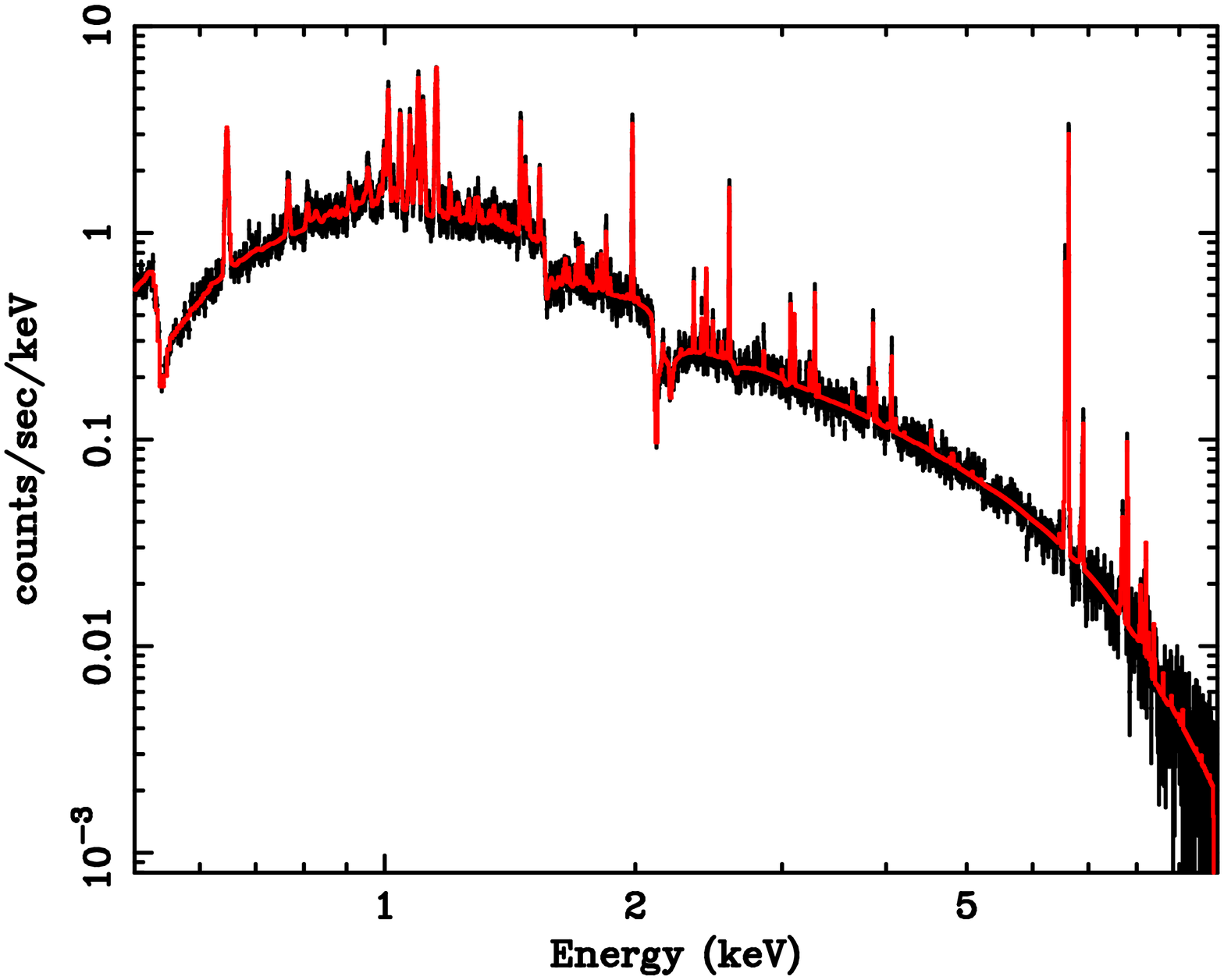}
\caption{Top panel: ASCA spectra of the Centaurus cluster.
The upper spectrum with circles is of GIS, and the lower spectrum is of SIS.
Bottom panel: A simulated XRS spectrum of the Centaurus cluster.} 
\label{cen}
\end{figure}


{\it Chandra} and {\it XMM-Newton} have brought us new views with
their ultimate imaging and spectroscopic capabilities. Astro-E2 is
ongoing to join them to make a substantial contribution to
science. The broad energy band-pass, spectroscopic resolving power
covering up to 10 keV, and applicability to extended sources will
provide a unique capability for dealing with fundamental problems in
astrophysics. Galaxy clusters with a thermal spectrum of $kT=$3--10
keV are the most suitable and interesting objects with XRS. We
face on the new features such as lack of cool gas below 1 keV, complex
cores together with radio lobes, various structures as a result of
subcluster mergers, and high metallicity rings.  The XRS data will
provide us a new, straightforward approach to solve those dynamical
and thermal problems in clusters.

In this section, we give four examples of scientific topics in galaxy
clusters based on XRS spectral simulations\footnote{Responses
used in this paper are prepared for the 4th Astro-E2 science working
group meeting held in November, 2003} to examine what kind of analysis
we will be able to do. Needless to say, there are much more other
interesting sciences that XRS possibly can do, for instance, thermal
broadening, and soft emission excess using Oxygen lines.

\subsection{Abundances of various metals (line intensity)}

First, we can divide complex lines into each single line so that we
will determine the line intensity more precisely ever before.  The top panel of
Figure \ref{cen} shows spectra of the Centaurus cluster taken by SIS
(CCD camera) and GIS (gas scintillation proportional counter) onboard ASCA
\citep{ikebe}. An apparent single line in the spectra is actually a
blend of several lines, which we can not divide with a CCD's
spectroscopic capability so far. However, 
we will finally resolve almost all lines of various
metals in interclutser and intergalactic media. The bottom panel of
Figure \ref{cen} is a simulated XRS spectrum of Centaurus, which shows
a lot of sharp emission lines. We will measure absolute abundances of
those lines from Oxygen to Nickel accurately and
model-independently. We will also obtain a spatial distribution of
each metal abundance even though it will be rough more than $30''$ step
depending on the XRS pixel size and the XRT imaging capability.

\begin{figure} [h]
\centering
\includegraphics[clip=true,width=0.48\textwidth]{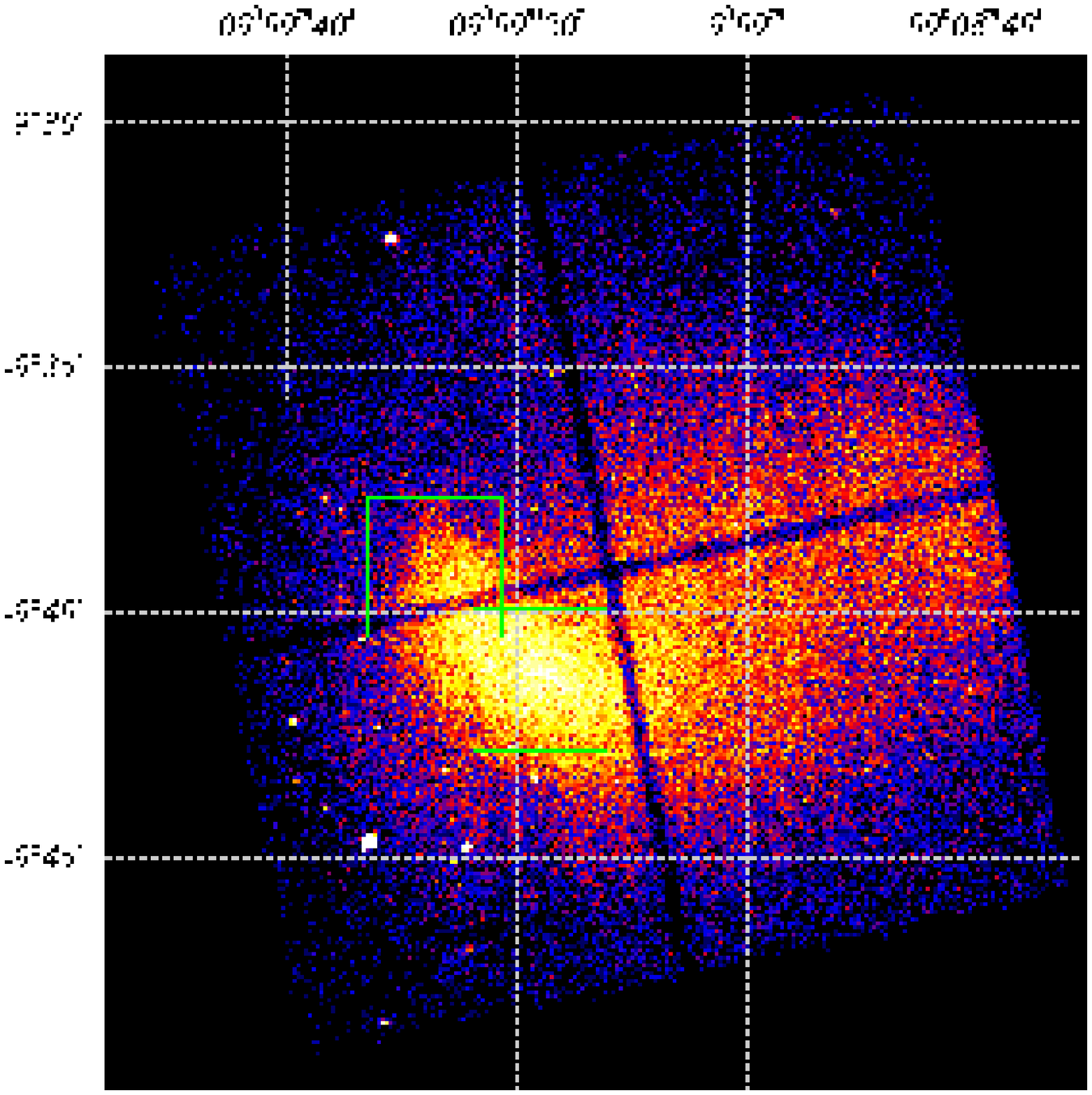}
\includegraphics[clip=true,angle=-90,width=0.48\textwidth]{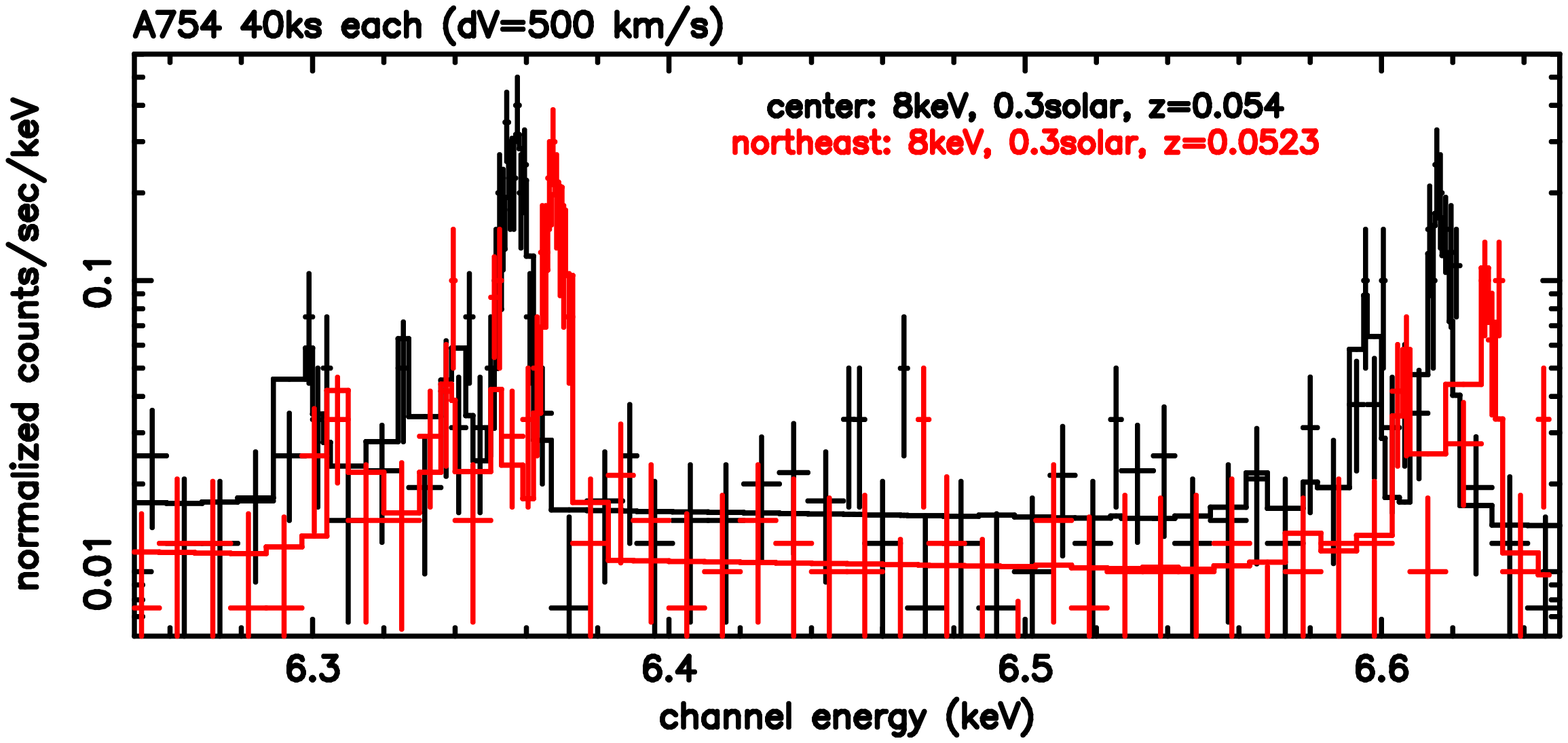}
	\caption{Top panel: Chandra ACIS-I image of Abell 754. 
The two green boxes represent the XRS field of view.
Bottom panel: Simulated XRS spectra of the two pointings overlayed on the 
image with 40 ks exposure each.}
   \label{a754} 
\end{figure}

\subsection{Gas motion (line shift)}

Second, we can determine line centers in accurate, and detect line
shifts using a Fe-K resonance line. The top panel of Figure \ref{a754}
is a Chandra image of Abell 754, which is a famous merging cluster.
ASCA and {\it Chandra} have shown that the cluster has a temperature
structure and a density gap that indicate an ongoing strong merger in
this cluster.  The intracluster gas where this kind of large merger
takes place expects that there is a gas bulk motion with a velocity of
several hundred to $\sim$1000 km/s.  The bottom panel shows simulated
XRS spectra for two pointing observations shown as two green boxes
in the top panel. We assume that an exposure time is 40 ks each, and a
line-of-sight difference of the gas velocity is 500 km/s. The velocity
difference, $\sim 10$ eV at 6 keV, is easily detected as a shift of
the line center. We can map out a velocity distribution in a direct way.

\begin{figure}[t] 
\centering
\vspace*{-1mm}
\hspace*{-1mm}
\includegraphics[angle=-90,width=0.51\textwidth]{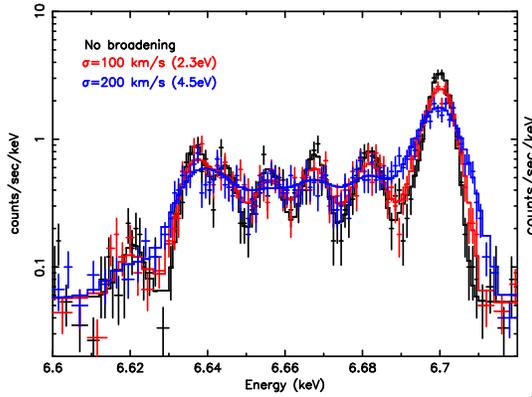}
\vspace*{-3mm}
\caption{Simulated spectra around Fe-K complex 
assuming with velocity dispersion by hydrodynamical turbulences of 
0, 100, and 200 km/s (black, red, and blue, respectively). }
   \label{turbl}
\end{figure}

\subsection{Turbulence (line broadening)}

Third, we can look into an existence of line broadening that means a
velocity dispersion by gas bulk motion.  Recently,
\citet{sunyaev}, and \citet{inogamo} reported that a line broadening by
hydrodynamic turbulences in the intergalactic gas could be present and
detected in many clusters. Emission lines of heavy elements,
especially of iron, are the most promising probes to measure turbulent
velocities, because their thermal broadening is much smaller compared
with that of protons. Figure \ref{turbl} shows simulated XRS spectra
for the complex of iron-K lines with an APEC model of $kT=3$ keV with
a velocity dispersion of 100 and 200 km/s and no line broadening. If
lines are broadened with a velocity of more than 200 km/s, they
create a flat-topped line profile that are immediately distinguishable 
from a spiky profile with no line broadening.

\subsection{Plasma diagnostics (line ratio)}

\begin{figure}[t] 
\centering
\vspace*{-1mm}
\hspace*{-2mm}
\includegraphics[angle=-90,width=0.51\textwidth]{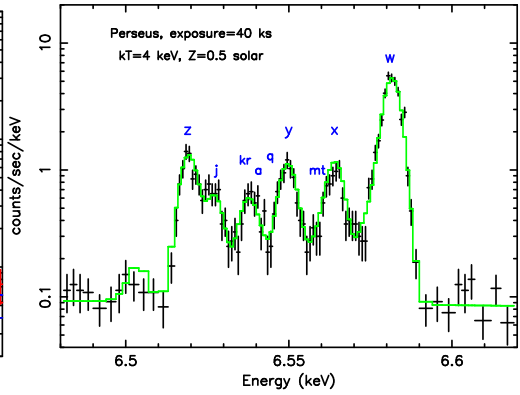}
\vspace*{-3mm}
 \caption{Fe XXV lines with a 40 ks simulation of Perseus cluster.}
   \label{fexxv}
\end{figure}

Forth, we can calculate line ratios of complex lines created by fine
structure.  We can separate a resonance line (w) and
inter-combination(x,y) and forbidden (z) lines, and possibly other
satellite lines (j,k,a etc.) as shown in Figure \ref{fexxv}. Line
ratios of those lines will be enable us to do plasma diagnostics with
Fe-K lines.  Plasma diagnostics is expected to give us important
physical parameters directly such as electron temperature, density,
and ionization level. Note that the line ratios also are useful of
direct analysis of resonance scattering in a dense cluster core such
as of M87.

\bigskip

Now, we proceed the final tests of each subsystem including
integration tests, calibration tests, performance tests, and
environmental tests. The PV phase targets of Astro-E2 will be
discussed soon and decided in March, 2004. First call for proposals in
the AO-1 phase, of which observations are starting six months later
after the launch, will be announced in next spring, 2004.

\begin{acknowledgements}
We would like to thank R.L. Kelley, 
K.R. Boyce, C.A. Kilbourne, J. Cottam, G.V. Brown, F.S. Porter, 
and everyone
working for Astro-E2/XRS at NASA/Goddard Space Flight Center, and
University of Wisconsin.  We also would like to thank all our members
of the Astro-E2 mission in Japan. T. F. is supported by the Japan
Society for the Promotion of Science (JSPS) Postdoctoral Fellowships.
\end{acknowledgements}

\bibliographystyle{aa}

\end{document}